\documentclass[final,british,conference]{IEEEtran}

\usepackage{cite}
\usepackage{graphicx}
\usepackage{amsmath}
\usepackage{array}
\usepackage{booktabs}
\usepackage{multirow}
\usepackage{float}
\usepackage{hyperref}
\usepackage{subfig}
\usepackage[font=small]{caption}
\usepackage{soul,color}
\newcolumntype{P}[1]{>{\centering\arraybackslash}p{#1}}
\newcolumntype{M}[1]{>{\centering\arraybackslash}m{#1}}
\newcommand\norm[1]{\left\lVert#1\right\rVert}
\usepackage[comma,sort&compress,numbers]{natbib}
\usepackage[linesnumbered,ruled,lined,commentsnumbered]{algorithm2e}
\usepackage{tabularx}

\begin{document}
%
\title{ViSTRA3: Video Coding with Deep Parameter Adaptation and Post Processing}

\author{\IEEEauthorblockN{Chen Feng\IEEEauthorrefmark{1}, Duolikun Danier\IEEEauthorrefmark{1}, Charlie Tan, Fan Zhang, and David Bull}
\IEEEauthorblockA{Visual Information Laboratory\\
University of Bristol\\
Bristol, BS8 1UB, United Kingdom \\
\{chen.feng, duolikun.danier, charlie.tan.2019, fan.zhang, dave.bull\}@bristol.ac.uk}
}


\maketitle
\begingroup\renewcommand\thefootnote{\IEEEauthorrefmark{1}}
\footnotetext{Chen and Danier contributed equally to this paper.}
\endgroup
\begin{abstract}
This paper presents a deep learning-based video compression framework (ViSTRA3). The proposed framework intelligently adapts video format parameters of the input video before encoding, subsequently employing a CNN at the decoder to restore their original format and enhance reconstruction quality. ViSTRA3 has been integrated with the H.266/VVC Test Model VTM 14.0, and evaluated under the Joint Video Exploration Team Common Test Conditions. Bjønegaard Delta (BD) measurement results show that the proposed framework consistently outperforms the original VVC VTM, with average BD-rate savings of 1.8\% and 3.7\% based on the assessment of PSNR and VMAF.
\end{abstract}


\IEEEpeerreviewmaketitle

\section{Introduction}

In recent years, there has been a significant increase in the amount of video consumed, associated with the demand for gaming, streaming, conferencing and broadcasting services \cite{bull2021intelligent}. Video constitutes the largest proportion of Internet traffic, which is likely to exceed 80\% by 2022~\cite{cisco2018cisco}. To alleviate the tension between the available bandwidth and the desired bit rate for video communication, there is an urgent requirement for new video compression algorithms with much higher coding efficiency. 

Conventional hybrid video coding standards have made remarkable progress in the past three decades. H.264/AVC~\cite{wiegand2003overview} was launched in 2003, revolutionising Internet-based video streaming, and remains the most widely deployed standard today. H.265/HEVC~\cite{sullivan2012overview} (2013) achieved significant coding gains (50\% on average) over H.264/AVC, while the most recent standard, H.266/VVC~\cite{bross2021overview} offers a further 40-50\% bit rate reduction compared to HEVC. 

Machine learning methods, in particular those based on convolutional neural networks (CNNs), have made a major impact across a wide range of computer vision and image processing applications~\cite{redmon2016you,he2016deep,9477504}. They have also been applied in video compression to enhance conventional coding tools~\cite{li2018fully,liu2018cnn,ma2020mfrnet,zhang2020enhancing} and to build new end-to-end coding architectures \cite{lu2020end, lin2020m, agustsson2020scale}. A further and related group of learning-based coding algorithms exploits the redundancy within a video format, adapting  spatial resolution, frame rate and bit depth according to content type and quality constraints. This CNN-based parameter restoration has been shown to provide significant coding gains over standard video codecs when integrated into different compression frameworks ~\cite{afonso2018video, zhang2021vistra2, jenab2018content, hosking2016adaptive}.

\begin{figure*}[t]
\centering
\includegraphics[width=\linewidth]{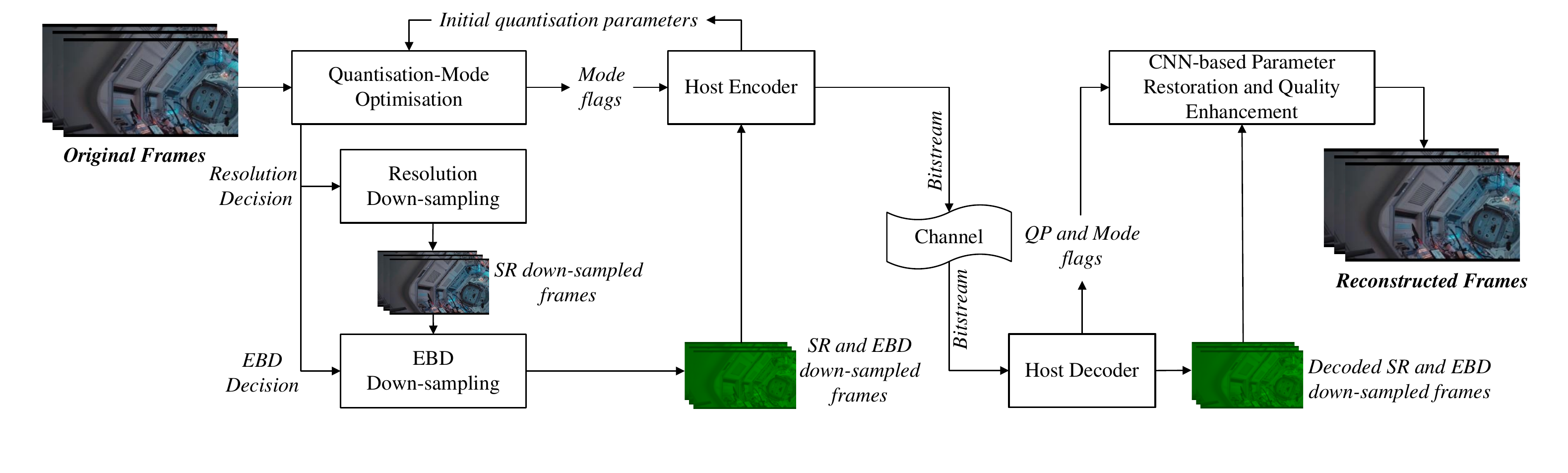}
\vspace{-8mm}
\caption{Diagram of the proposed coding framework.\label{fig:framework}}
\vspace{-5mm}
\end{figure*}

In this paper, we present a new coding framework, ViSTRA3 (Fig. \ref{fig:framework}), built upon our previous resolution adaptation and post-processing architectures~\cite{afonso2018video, zhang2021vistra2,zhang2021video}. ViSTRA3 adaptively predicts the optimal video parameters during encoding using a CNN-based Quantisation-Mode Optimisation (QMO) module. The original format is then restored at the decoder using an advanced CNN architecture, MFRNet~\cite{ma2020mfrnet}, which also enhances the final reconstruction quality. The BVI-DVC database~\cite{ma2021bvi} has been used as the training material to optimise all the CNN models in this coding framework. ViSTRA3 has been evaluated using the Joint Video Experts Team (JVET) Common Test Conditions, and the results show consistent bit rate savings when video quality is assessed by PSNR and VMAF~\cite{li2016toward}. The main contributions of this work are summarised below.

\begin{itemize}
    \item Development of a novel video coding system integrating four deep learning based coding modes including effective bit depth adaptation (EBDA), spatial resolution adaptation (SRA), spatial resolution and effective bit depth adaptation (SREBDA), and post-processing (PP).
    \item Employment of an efficient CNN for learning optimal parameter settings directly from the original input video frames.
    \item Use of an advanced CNN architecture, the Multi-level Feature Review Network (MFRNet), for video parameter restoration and quality enhancement.
\end{itemize}

The remainder of this paper is structured as follows. In Section~\ref{sec:framework}, the proposed coding framework is described in detail alongside the mode selection process and the parameter restoration/quality enhancement method. Section~\ref{sec:results} focuses on the evaluation results and provides a critical analysis of the coding performance. Finally, Section \ref{sec:conclusion} concludes the paper and suggests possibilities for future work.

\begin{figure}[t]
\centering
\includegraphics[width=\linewidth]{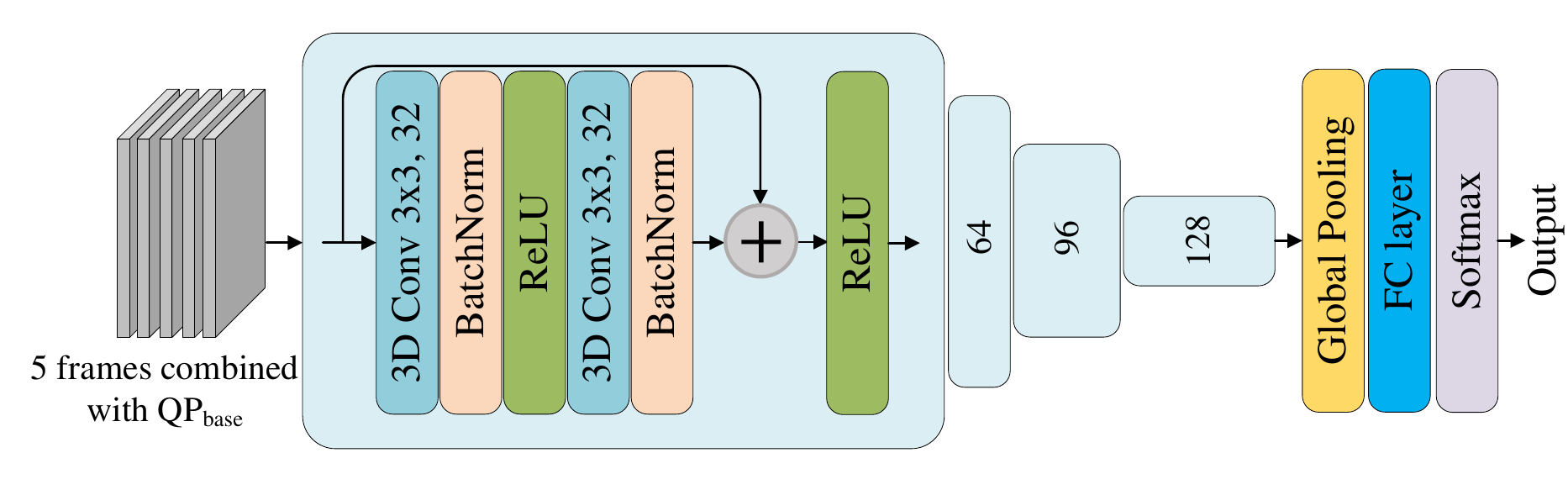}
\vspace{-5mm}
\caption{Architecture of the 3D CNN used for Quantisation-Mode Optimisation.\label{fig:modecnn}}
\vspace{-5mm}
\end{figure}
\section{Proposed Algorithm}\label{sec:framework}

The proposed coding framework is illustrated in Fig.~\ref{fig:framework}. Given original video content and an initial quantisation parameter ($\mathrm{QP}_\mathrm{base}$), the Quantisation-Mode-Optimisation (QMO) module selects the optimum adaptation mode from four  candidates (M0-M4): 

\begin{enumerate}
    \item[M0)] \textbf{Default}: no parameter adaptation or post-processing is invoked. Only the host codec is used for coding.

    \item[M1)] \textbf{Effective bit depth adaptation (EBDA)}: the effective bit depth (EBD) of the input video frame is down-sampled by 1 bit using bitwise-shifting for both luma and chroma channels. Here EBD refers to the actual bit depth used to represent the video signal. This differs from the pixel bit depth for encoding (e.g. ``InternalBitDepth" in HEVC and VVC Test Models). The latter is kept constant in the proposed coding workflow. More information on EBD can be found in \cite{EBDA,zhang2021vistra2}.
    
    \item[M2)] \textbf{Spatial resolution adaptation (SRA)}: the spatial resolution of the input video frame is down-sampled by a factor of two using a Lanczos3 filter~\cite{Lanczos3}.
    
    \item[M3)] \textbf{Spatial resolution and effective bit depth adaptation (SREBDA)}: both  spatial resolution and EBD are down-sampled using the methods described for M1 and M2.
    
    \item[M4)] \textbf{Post-processing (PP)}: all video format parameters are unaltered; only post processing is invoked at the decoder.
\end{enumerate}

The mode decision is indicated using a one-byte flag, which is included in the bitstream as side information. The pre-processed frames are then passed to the host encoder, which performs video compression using a modified $\mathrm{QP}=\mathrm{QP}_\mathrm{base}-\Delta\mathrm{QP}$, where $\Delta\mathrm{QP}$ is an offset applied according to the adaptation mode decided by the QMO. This offset is designed to produce similar bit rates to those generated by the original host encoder (using $\mathrm{QP}_\mathrm{base}$ without adaptation). $\Delta\mathrm{QP}$ is set to -6 for modes M1 and M2, -12 for M3, and 0 for M0 and M4. These QP offset values have been determined empirically by examining the coding statistics of a wide range of training sequences \cite{SRA,EBDA}. 

During decoding, the ViSTRA3 decoder first retrieves the mode and QP information from the bitstream, and then reconstructs each video frame using the adapted video format (or with the original video format for M0 and M4). Depending on the adaptation mode invoked, each video frame is then processed by a CNN to restore its original format (or just to enhance final quality for M4).

The details of the QMO module, the CNN architecture and the training process are described below.

\subsection{Quantisation-Mode Optimisation}

It has been previously reported in~\cite{hosking2016adaptive,afonso2018video,zhang2021vistra2} that the effectiveness of the different parameter adaptation methods is highly content-dependent. Hand-crafted features were employed in these works to optimise the rate quality performance. In contrast, in ViSTRA3, we introduce a Quantisation-Mode Optimisation (QMO) module to determine the optimal parameter adaptation mode, taking account of input content characteristics and quantisation level. 

In the QMO module, a CNN is designed to take five consecutive original video frames and the base quantisation parameter $\mathrm{QP}_\mathrm{base}$ as the input, and to output a mode decision (M0, M1, M2, M3 or M4). Inspired by the coord-conv trick~\cite{bailer2015flow, cheng2021multiple}, we create a 2D matrix, each entry of which has the same value as $\mathrm{QP}_\mathrm{base}$. This is then concatenated with each input frame as an additional channel. The resulting five 4-channel frames are then processed using the CNN. As shown in Fig.~\ref{fig:modecnn}, the CNN architecture used here is a simplified version of the 3D ResNet~\cite{tran2018closer}, which comprises a series of 3D convolution and ReLU activation layers. In the output stage, the intermediate 3D features are then pooled and processed by the final fully-connected layer, which outputs the mode decision using a softmax function.

\begin{figure*}[t]
\centering
\includegraphics[width=\linewidth]{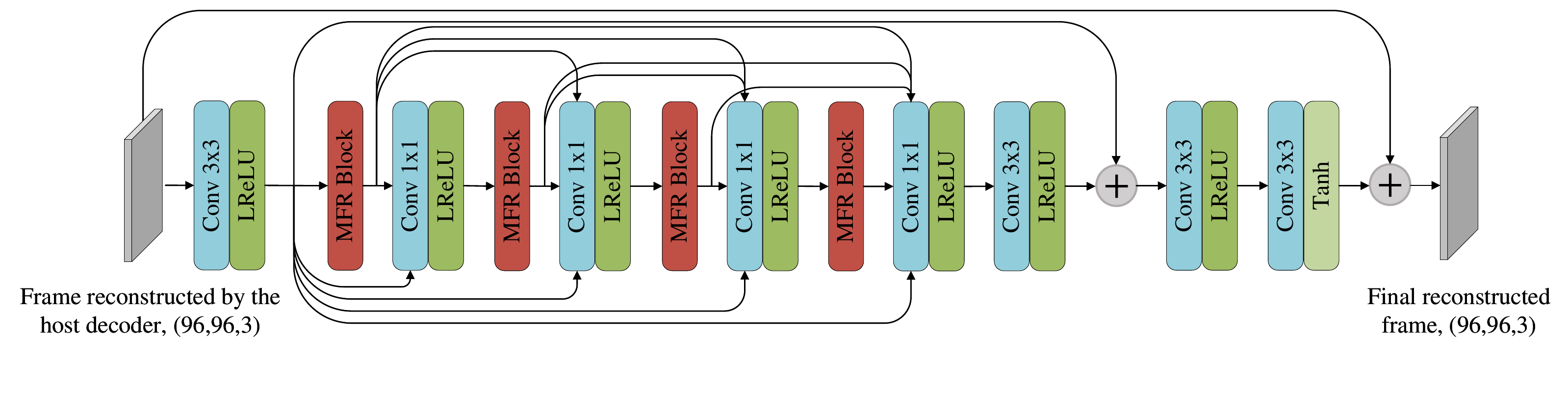}
\vspace{-10mm}
\caption{Architecture of the MFRNet for parameter re-sampling. Figure adapted from~\cite{ma2020mfrnet}.\label{fig:mfrnet}}
\vspace{-3mm}
\end{figure*}

\begin{algorithm}[t]
\small
    \SetKwInOut{Input}{Input}
    \SetKwInOut{Output}{Output}
	\SetAlgoLined

	\Input{HD sequence $\mathcal{S}$ containing 64 frames}
	\Input{Quantisation parameters $\mathcal{Q}=\{22,27,32,37\}$}
	\Input{Adaptation mode $\mathcal{M}=\{\text{M0,M1,M2,M3,M4}\}$}
	\Output{Training instances $\mathcal{D}=\{(\mathcal{X}_k, \mathcal{Y}_k)_{k=1\dots 2560}\}$}
	\BlankLine
	
	Initialise empty collection $\mathcal{D}$ \\
	\For{$i=1$ to $64$}{
	    Randomly crop a 32-frame sequence $\mathcal{S}_{tmp}$ of size 256$\times$256 from $\mathcal{S}$ \\
	    Encode and decode $\mathcal{S}_\mathrm{tmp}$ with VTM 14.0 at each $\mathrm{QP}_\mathrm{base}\in \mathcal{Q}$ and obtain an anchor rate-PSNR curve via cubic interpolation\\
	    \ForEach{$\mathrm{QP}_\mathrm{base}\in \mathcal{Q}$}{
	        \ForEach{$M \in \mathcal{M}$}{
	            Get $\mathrm{QP}$ (w/ offset) based on $\mathrm{QP}_\mathrm{base}$ and $M$ \\
	            Encode and Decode $\mathcal{S}_\mathrm{tmp}$ using mode $M$ at $\mathrm{QP}$ and get a rate-PSNR point\\
	        }
	        Compare the 4 rate-PSNR points to the anchor rate-PSNR curve and get the best mode $M_\mathrm{best}$\\
	        Set the ground-truth optimal mode $\mathcal{Y}_\mathrm{tmp}=M_\mathrm{best}$ for the pair $(\mathcal{S}_{tmp}, \mathrm{QP}_\mathrm{base})$ \\
	        \For{$j=1$ to $10$}{
                Randomly crop 5-frame sequence $\mathcal{S}_j$ from $\mathcal{S}_\mathrm{tmp}$ \\
                Set $\mathcal{X}_j=(\mathcal{S}_j,\mathrm{QP}_\mathrm{base})$, $\mathcal{Y}_j=\mathcal{Y}_\mathrm{tmp}$ \\
                Append $(\mathcal{X}_j, \mathcal{Y}_j)$ to $\mathcal{D}$
	        }
	    }
	}
	\caption{\small Training Data Generation for QMO} \label{alg:qmodata}
\end{algorithm}

We have trained this CNN based on the methodology which has been previous employed in \cite{zhang2021vistra2}, using 200 HD source sequences from the BVI-DVC~\cite{ma2021bvi} database. Algorithm \ref{alg:qmodata} describes the training instance generation process for each source sequence in detail, and this produces a total of 512,000 $(\mathcal{X}_j, \mathcal{Y}_j)$ pairs for training the CNN. The target ground truth $\{\mathcal{Y}_j\}$ was then converted to one-hot encoded vectors during the training process. The CNN was then trained using cross-entropy loss, formulating mode selection as a multi-class classification problem. 

At the evaluation stage, every 5 consecutive frames are concatenated with the $\mathrm{QP}_\mathrm{base}$ and passed to the CNN to predict an optimal mode. When there is a change of predicted optimal modes during the coding process, we adopted the video segmentation approach similar to {\cite{ViSTRA2}}, where videos are truncated into multiple segments, and each segment with the same optimal mode is encoded separately with an individual mode flag. The flag uses one byte to represent M0, M1, M2, M3 or M4. In order to avoid excessive segmentation (which may lead to performance loss) two criteria are employed for a segmentation:

\begin{itemize}
    \item Each segment should be at least 1 second in duration.
    \item The predicted probability (the output of the CNN softmax layer) of the optimal mode must be at least 70\%.
\end{itemize}

These criteria constrain the extent of segmentation, maintaining the distinct characteristics of each segment, and do not introduce extra delay if a Random Access configuration is used.

\subsection{Parameter Restoration Network}

For all four adaptation modes other than M0, the reconstructed frames produced by the host decoder are fed into a deep CNN to perform video format restoration (for M1, M2 or M3) or video quality enhancement (for M4). In ViSTRA3, we employ the Multi-level Feature Review Network (MFRNet)~\cite{ma2020mfrnet}, which was originally developed for post-processing and in-loop filtering. MFRNet contains multiple Multi-level Feature Review blocks (MFRBs), each of which processes both the main output and and high-dimensional features produced by previous blocks. With numerous carefully designed residual connections, MFRNet has been reported to offer greater coding gains compared to other network architectures when used for post-processing and in-loop filtering. Further details on MFRNet can be found in~\cite{ma2020mfrnet}.

\subsection{Network Training and Evaluation}

The training data for the format restoration network was generated from the 200 HD sequences of the BVI-DVC dataset~\cite{ma2021bvi}. For each adaptation mode, all original sequences were encoded using the host encoder (VVC VTM 14.0) at four $\mathrm{QP}_\mathrm{base}$ (22, 27, 32, 37), with the corresponding offset configured for M0, M1, M2, M3 or M4. The Random Access (RA) mode in the Joint Video Exploration Team (JVET) Common Test Conditions (CTC)~\cite{bossen2019jvet} was used.

This resulted in 16 groups (4 adaptation modes $\times$ 4 QPs) of reconstructed videos, where each group contained 200 compressed sequences. For each group, around 50,000 96$\times$96 patches together with their ground truth (original uncompressed patches) counterparts were cropped and selected randomly (with augmentation based on rotation and flips).

All CNN models were trained using a combination of the Laplacian Pyramid loss~\cite{bojanowski2017optimizing} and $\mathcal{L}_1$ loss:
\begin{equation}
    \mathcal{L} = 10\mathcal{L}_{lap} + \mathcal{L}_1 
    \end{equation}
in which 
\begin{equation}
    \left\{
    \begin{array}{ll}
    \mathcal{L}_{lap} = \sum_{s=1}^{S} 2^{s-1} \norm{L^s(I_{out})-L^s(I_{gt})}_1 &\\
    \mathcal{L}_1 = \norm{I_{out} - I_{gt}}_1&
    \end{array}
    \right.
\end{equation}
where $I_{out},I_{gt}$ are the network output and ground-truth images respectively, $L^s(\cdot)$ denotes the $s$th level of the Laplacian pyramid of an image with $S$ being the maximum level. The optimizer used was ADAM~\cite{kingma2014adam} with $\beta_1$=0.9 and $\beta_2$=0.999. The network was trained using a batch size of 16 for 50 epochs with the initial learning rate of 0.0001 halved every 20 epochs. All the source code was implemented in TensorFlow~\cite{tensorflow}. A total of 16 CNNs have been trained and, in the evaluation stage, the corresponding CNN model for the given adaptation mode and $\mathrm{QP}_\mathrm{base}$ is invoked. All training and evaluation were performed on a shared cluster, BlueCrystal~\cite{BC4}. Specifically, the computation nodes contain 2.4GHz Intel CPUs and NVIDIA P100 GPUs.

\section{Results and Discussion}\label{sec:results}

 Our integrated coding framework has been evaluated on the five HD sequences in JVET CTC  for the evaluation experiment. We have included two anchor codecs for benchmarking the proposed coding algorithm: HEVC HM 16.22 and VVC VTM 14.0. While VTM was used as an anchor because it is the host codec of ViSTRA3, HM was also compared against for reference. During evaluation, the Random Access configuration was employed for both codecs with four $\mathrm{QP}_\mathrm{base}$ values 22, 27, 32 and 37. Different offsets were applied in ViSTRA3 according to the selected adaptation mode (as described in Section \ref{sec:framework}).

\begin{table}[t]
\centering

\caption{Compression results of the proposed framework integrated with VTM 14.0 in terms of PSNR.}

\begin{tabularx}{\linewidth}{l||>{\centering\arraybackslash}X|>{\centering\arraybackslash}c||>{\centering\arraybackslash}X|>{\centering\arraybackslash}c}
\toprule
Anchor &  \multicolumn{2}{c||}{HM 16.22} & 
\multicolumn{2}{c}{VTM 14.0}\\
\cmidrule{1-5}
\centering
Sequence & BD-rate & BD-PSNR & BD-rate & BD-PSNR\\
\midrule \midrule
BasketballDrive &-45.5\%& +1.09dB &-2.8\%& +0.04dB\\
\midrule
BQTerrace &-59.8\%& +0.95dB&-0.7\%& +0.01dB\\
\midrule
Cactus &-51.8\%& +1.20dB&-2.0\%& +0.03dB\\
\midrule
MarketPlace &-45.8\%& +1.31dB&-2.2\%& +0.04dB\\
\midrule
RitualDance &-37.2\%& +1.77dB&-1.4\%& +0.05dB\\
\midrule \midrule 
\textbf{Overall} &-48.0\%& +1.26dB&-1.8\%&  +0.03dB\\
\bottomrule
\end{tabularx}

\label{tab:res1}
\end{table}
\begin{table}[t]
\centering
\caption{Compression results of the proposed framework integrated with VTM 14.0 in terms of VMAF.}

\begin{tabularx}{\linewidth}{l||>{\centering\arraybackslash}X|>{\centering\arraybackslash}c||>{\centering\arraybackslash}X|>{\centering\arraybackslash}c}
\toprule
Anchor &  \multicolumn{2}{c||}{HM 16.22} & 
\multicolumn{2}{c}{VTM 14.0}\\
\cmidrule{1-5}
\centering
Sequence & BD-rate & BD-VMAF & BD-rate & BD-VMAF\\
\midrule \midrule
BasketballDrive &-45.6\%& +4.34&-3.5\%& +0.18 \\
\midrule
BQTerrace &-64.7\%&+2.33 &-1.2\%& +0.00\\
\midrule
Cactus &-53.8\%&+5.46 &-4.8\%& +0.29\\
\midrule
MarketPlace &-50.3\%&+6.94 &-4.7\%& +0.38\\
\midrule
RitualDance &-39.2\%& +5.64&-4.9\%& +0.47\\
\midrule \midrule 
\textbf{Overall} &-50.7\%& +4.94 &-3.8\%&+0.26  \\
\bottomrule
\end{tabularx}

\label{tab:res2}
\end{table}
\begin{table}[t]
\centering
\caption{Computational Complexity of  ViSTRA3.}
\begin{tabularx}{\linewidth}{l|>{\centering\arraybackslash}X|>{\centering\arraybackslash}X|>{\centering\arraybackslash}X|>{\centering\arraybackslash}X}
\toprule
\multirow{2}{*}{Anchor}	& \multicolumn{2}{c|}{HM 16.22} & \multicolumn{2}{c}{VTM 14.0}	\\ 
\cmidrule{2-5}
	  & Encoder & Decoder & Encoder & Decoder\\
\midrule Average &   1004\% & 9630\% & 102\% & 6485\%\\
\bottomrule	
\end{tabularx}
\label{tab:complexity}
\vspace{-5mm}
\end{table}

The rate-quality performance of ViSTRA3 is compared to anchor codecs based on Peak Signal-to-Noise Ratio (PSNR) and the Bj\o ntegaard delta measurements (BD-rate and BD-PSNR)~\cite{bjontegaard2001calculation}. Specifically, PSNR is calculated on all Y, U, V channels using Equation~(\ref{eqn:psnr}):
\begin{equation}
    \text{PSNR}_{YUV} = (6\times\text{PSNR}_Y+\text{PSNR}_U+\text{PSNR}_V)/8 \label{eqn:psnr}
\end{equation}
To further provide more accurate prediction of perceptual quality \cite{zhang2018bvi}, an additional quality metric, Video Multimethod Assessment fusion (VMAF) \cite{li2016toward}, has been employed here. 

Table~\ref{tab:res1} and~\ref{tab:res2} summarises the coding performance of our proposed framework against the two anchor codecs on the five HD test sequences in JVET-CTC. It can be observed that ViSTRA3 offers significant coding gains over HM 16.22, with average BD-rates of -48.0\% and -50.7\% for PSNR and VMAF respectively. When VTM is used as the benchmark, ViSTRA3 again achieves consistent bit rate savings on all evaluated sequences, with overall BD-rate figures of -1.8\% and -3.8\% for these two metrics.

The complexity of the proposed framework was also compared against the two anchor codecs, with results shown in Table~\ref{tab:complexity}. The decoding complexity of ViSTRA3 is much higher than the original HM and VTM due to the use of CNN for format restoration and quality enhancement. The encoder complexity is similar to VTM and higher than HM, and this is because of the employment of VTM 14.0 as the host encoder.

\section{Conclusion}
\label{sec:conclusion}
In this paper, a new video compression framework, ViSTRA3, has been proposed which exploits parameter adaptation and post-processing. ViSTRA3 supports intelligent mode optimisation, which maximises overall rate-quality performance. Mode optimisation is performed using a efficient 3D CNN which extracts spatio-temporal features from the input video and then predicts the optimal video parameters for encoding. At the decoder, an advanced CNN architecture, MFRNet, was employed to restore original parameters and further improve video quality. Quantitative results on the JVET-CTC test sequences show that ViSTRA3 achieves 48.0\% and 1.8\% overall coding gains (based on PSNR) over HEVC and VVC anchors respectively.

\section*{Acknowledgment}

This work was funded by the University of Bristol, the MyWorld Strength in Places Programme and the China Scholarship Council (Duolikun Danier).

{
\small
\bibliographystyle{IEEEtran}
\bibliography{ref}

\begin{thebibliography}{10}
\providecommand{\url}[1]{#1}
\csname url@samestyle\endcsname
\providecommand{\newblock}{\relax}
\providecommand{\bibinfo}[2]{#2}
\providecommand{\BIBentrySTDinterwordspacing}{\spaceskip=0pt\relax}
\providecommand{\BIBentryALTinterwordstretchfactor}{4}
\providecommand{\BIBentryALTinterwordspacing}{\spaceskip=\fontdimen2\font plus
\BIBentryALTinterwordstretchfactor\fontdimen3\font minus
  \fontdimen4\font\relax}
\providecommand{\BIBforeignlanguage}[2]{{%
\expandafter\ifx\csname l@#1\endcsname\relax
\typeout{** WARNING: IEEEtran.bst: No hyphenation pattern has been}%
\typeout{** loaded for the language `#1'. Using the pattern for}%
\typeout{** the default language instead.}%
\else
\language=\csname l@#1\endcsname
\fi
#2}}
\providecommand{\BIBdecl}{\relax}
\BIBdecl

\bibitem{bull2021intelligent}
D.~R. Bull and F.~Zhang, \emph{Intelligent image and video compression:
  communicating pictures}.\hskip 1em plus 0.5em minus 0.4em\relax Academic
  Press, 2021.

\bibitem{cisco2018cisco}
{CISCO, VNI}, ``{CISCO} visual networking index: Forecast and trends,
  2017--2022 white paper.'' 2018.

\bibitem{wiegand2003overview}
T.~Wiegand, G.~J. Sullivan, G.~Bj{\o}ntegaard, and A.~Luthra, ``Overview of the
  {H.264/AVC} video coding standard,'' \emph{IEEE Transactions on circuits and
  systems for video technology}, vol.~13, no.~7, pp. 560--576, 2003.

\bibitem{sullivan2012overview}
G.~J. Sullivan, J.-R. Ohm, W.-J. Han, and T.~Wiegand, ``Overview of the high
  efficiency video coding ({HEVC}) standard,'' \emph{IEEE Transactions on
  circuits and systems for video technology}, vol.~22, no.~12, pp. 1649--1668,
  2012.

\bibitem{bross2021overview}
B.~Bross, Y.-K. Wang, Y.~Ye, S.~Liu, J.~Chen, G.~J. Sullivan, and J.-R. Ohm,
  ``Overview of the versatile video coding ({VVC}) standard and its
  applications,'' \emph{IEEE Transactions on Circuits and Systems for Video
  Technology}, vol.~31, no.~10, pp. 3736--3764, 2021.

\bibitem{redmon2016you}
J.~Redmon, S.~Divvala, R.~Girshick, and A.~Farhadi, ``You only look once:
  Unified, real-time object detection,'' in \emph{Proceedings of the IEEE
  conference on computer vision and pattern recognition}, 2016, pp. 779--788.

\bibitem{he2016deep}
K.~He, X.~Zhang, S.~Ren, and J.~Sun, ``Deep residual learning for image
  recognition,'' in \emph{Proceedings of the IEEE conference on computer vision
  and pattern recognition}, 2016, pp. 770--778.

\bibitem{9477504}
D.~Danier and D.~Bull, ``Texture-aware video frame interpolation,'' in
  \emph{2021 Picture Coding Symposium (PCS)}, 2021, pp. 1--5.

\bibitem{li2018fully}
J.~Li, B.~Li, J.~Xu, R.~Xiong, and W.~Gao, ``Fully connected network-based
  intra prediction for image coding,'' \emph{IEEE Transactions on Image
  Processing}, vol.~27, no.~7, pp. 3236--3247, 2018.

\bibitem{liu2018cnn}
D.~Liu, H.~Ma, Z.~Xiong, and F.~Wu, ``{CNN-based DCT}-like transform for image
  compression,'' in \emph{International Conference on Multimedia
  Modeling}.\hskip 1em plus 0.5em minus 0.4em\relax Springer, 2018, pp. 61--72.

\bibitem{ma2020mfrnet}
D.~Ma, F.~Zhang, and D.~R. Bull, ``{MFRNet}: a new cnn architecture for
  post-processing and in-loop filtering,'' \emph{IEEE Journal of Selected
  Topics in Signal Processing}, vol.~15, no.~2, pp. 378--387, 2020.

\bibitem{zhang2020enhancing}
F.~Zhang, C.~Feng, and D.~R. Bull, ``Enhancing {VVC} through {CNN-based}
  post-processing,'' in \emph{2020 IEEE International Conference on Multimedia
  and Expo (ICME)}.\hskip 1em plus 0.5em minus 0.4em\relax IEEE, 2020, pp.
  1--6.

\bibitem{lu2020end}
G.~Lu, X.~Zhang, W.~Ouyang, L.~Chen, Z.~Gao, and D.~Xu, ``An end-to-end
  learning framework for video compression,'' \emph{IEEE transactions on
  pattern analysis and machine intelligence}, 2020.

\bibitem{lin2020m}
J.~Lin, D.~Liu, H.~Li, and F.~Wu, ``{M-LVC}: multiple frames prediction for
  learned video compression,'' in \emph{Proceedings of the IEEE/CVF Conference
  on Computer Vision and Pattern Recognition}, 2020, pp. 3546--3554.

\bibitem{agustsson2020scale}
E.~Agustsson, D.~Minnen, N.~Johnston, J.~Balle, S.~J. Hwang, and G.~Toderici,
  ``Scale-space flow for end-to-end optimized video compression,'' in
  \emph{Proceedings of the IEEE/CVF Conference on Computer Vision and Pattern
  Recognition}, 2020, pp. 8503--8512.

\bibitem{afonso2018video}
M.~Afonso, F.~Zhang, and D.~R. Bull, ``Video compression based on
  spatio-temporal resolution adaptation,'' \emph{IEEE Transactions on Circuits
  and Systems for Video Technology}, vol.~29, no.~1, pp. 275--280, 2018.

\bibitem{zhang2021vistra2}
F.~Zhang, M.~Afonso, and D.~R. Bull, ``{ViSTRA2}: Video coding using spatial
  resolution and effective bit depth adaptation,'' \emph{Signal Processing:
  Image Communication}, p. 116355, 2021.

\bibitem{jenab2018content}
M.~Jenab, I.~Amer, B.~Ivanovic, M.~Saeedi, Y.~Liu, G.~Sines, and S.~Shirani,
  ``Content-adaptive resolution control to improve video coding efficiency,''
  in \emph{2018 IEEE International Conference on Multimedia \& Expo Workshops
  (ICMEW)}.\hskip 1em plus 0.5em minus 0.4em\relax IEEE, 2018, pp. 1--4.

\bibitem{hosking2016adaptive}
B.~Hosking, D.~Agrafiotis, D.~Bull, and N.~Eastern, ``An adaptive resolution
  rate control method for intra coding in {HEVC},'' in \emph{2016 IEEE
  International Conference on Acoustics, Speech and Signal Processing
  (ICASSP)}.\hskip 1em plus 0.5em minus 0.4em\relax IEEE, 2016, pp. 1486--1490.

\bibitem{zhang2021video}
F.~Zhang, D.~Ma, C.~Feng, and D.~R. Bull, ``Video compression with {CNN-based}
  post processing,'' \emph{IEEE MultiMedia}, 2021.

\bibitem{ma2021bvi}
D.~Ma, F.~Zhang, and D.~Bull, ``{BVI-DVC}: a training database for deep video
  compression,'' \emph{IEEE Transactions on Multimedia}, 2021.

\bibitem{li2016toward}
Z.~Li, A.~Aaron, I.~Katsavounidis, A.~Moorthy, and M.~Manohara, ``Toward a
  practical perceptual video quality metric,'' \emph{The Netflix Tech Blog},
  vol.~6, no.~2, 2016.

\bibitem{EBDA}
F.~Zhang, M.~Afonso, and D.~R. Bull, ``Enhanced video compression based on
  effective bit depth adaptation,'' in \emph{2019 IEEE International Conference
  on Image Processing (ICIP)}, 2019, pp. 1720--1724.

\bibitem{Lanczos3}
K.~Turkowski, \emph{Filters for Common Resampling Tasks}.\hskip 1em plus 0.5em
  minus 0.4em\relax USA: Academic Press Professional, Inc., 1990, p. 147–165.

\bibitem{SRA}
M.~Afonso, F.~Zhang, A.~Katsenou, D.~Agrafiotis, and D.~Bull, ``Low complexity
  video coding based on spatial resolution adaptation,'' in \emph{2017 IEEE
  International Conference on Image Processing (ICIP)}, 2017, pp. 3011--3015.

\bibitem{bailer2015flow}
C.~Bailer, B.~Taetz, and D.~Stricker, ``Flow fields: Dense correspondence
  fields for highly accurate large displacement optical flow estimation,'' in
  \emph{Proceedings of the IEEE international conference on computer vision},
  2015, pp. 4015--4023.

\bibitem{cheng2021multiple}
X.~Cheng and Z.~Chen, ``Multiple video frame interpolation via enhanced
  deformable separable convolution,'' \emph{IEEE Transactions on Pattern
  Analysis and Machine Intelligence}, 2021.

\bibitem{tran2018closer}
D.~Tran, H.~Wang, L.~Torresani, J.~Ray, Y.~LeCun, and M.~Paluri, ``A closer
  look at spatiotemporal convolutions for action recognition,'' in
  \emph{Proceedings of the IEEE conference on Computer Vision and Pattern
  Recognition}, 2018, pp. 6450--6459.

\bibitem{ViSTRA2}
F.~Zhang, M.~Afonso, and D.~R. Bull, ``{ViSTRA2}: Video coding using spatial
  resolution and effective bit depth adaptation,'' \emph{Signal Processing:
  Image Communication}, vol.~97, p. 116355, Sep 2021.

\bibitem{bossen2019jvet}
F.~Bossen, J.~Boyce, X.~Li, V.~Seregin, and K.~S{\"u}hring, ``{JVET} common
  test conditions and software reference configurations for sdr video,''
  \emph{Joint Video Experts Team (JVET) of ITU-T SG}, vol.~16, pp. 19--27,
  2019.

\bibitem{bojanowski2017optimizing}
P.~Bojanowski, A.~Joulin, D.~Lopez-Paz, and A.~Szlam, ``Optimizing the latent
  space of generative networks,'' \emph{arXiv preprint arXiv:1707.05776}, 2017.

\bibitem{kingma2014adam}
D.~P. Kingma and J.~Ba, ``Adam: a method for stochastic optimization,''
  \emph{arXiv preprint arXiv:1412.6980}, 2014.

\bibitem{tensorflow}
M.~Abadi, A.~Agarwal, P.~Barham, E.~Brevdo, Z.~Chen, C.~Citro, G.~S. Corrado,
  A.~Davis, J.~Dean, M.~Devin \emph{et~al.}, ``Tensorflow: Large-scale machine
  learning on heterogeneous distributed systems,'' \emph{arXiv preprint
  arXiv:1603.04467}, 2016.

\bibitem{BC4}
{University of Bristol}, ``Bluecystal phase 4,''
  http://www.acrc.bris.ac.uk/acrc/phase4.htm, accessed: 1st May 2017.

\bibitem{bjontegaard2001calculation}
G.~Bj{\o}ntegaard, ``Calculation of average {PSNR} differences between
  rd-curves,'' \emph{VCEG-M33}, 2001.

\bibitem{zhang2018bvi}
F.~Zhang, F.~M. Moss, R.~Baddeley, and D.~R. Bull, ``{BVI-HD}: A video quality
  database for hevc compressed and texture synthesized content,'' \emph{IEEE
  Transactions on Multimedia}, vol.~20, no.~10, pp. 2620--2630, 2018.

\end{thebibliography}
}

\end{document}